\documentclass[12pt]{article}
	\usepackage{latexsym,graphicx,multirow}
	\usepackage{amssymb}
	\usepackage{amsmath}
	\usepackage{amscd}
	\usepackage{amsthm}
	\usepackage[left=2cm,top=2.5cm,right=2.5cm,bottom=1.5cm]{geometry}
	\usepackage{hyperref}
	\usepackage{epstopdf}
	\usepackage{float}
	\usepackage{subcaption}
	
\begin{document}

	\begin{center}
	\large{\bf{Evaluation of cosmological models in $f(R, T)$ gravity in different dark energy scenario}} \\
	\vspace{10mm}
	\normalsize{ Vinod Kumar Bhardwaj$^1$, Anirudh Pradhan$^2$ }\\
	\vspace{5mm}
	\normalsize{$^{1,2}$Department of Mathematics, Institute of Applied Sciences and Humanities, GLA University,\\
		Mathura-281 406, Uttar Pradesh, India}\\
	\vspace{2mm}
	$^1$E-mail: dr.vinodbhardwaj@gmail.com\\
	$^2$E-mail: pradhan.anirudh@gmail.com\\

	\vspace{10mm}
	
	\end{center}
        \begin{abstract}
	
	In present paper, we search the existence of dark energy scalar field models within in $f(R, T)$ gravity 
	theory established by Harko et al. (Phys. Rev. D 84, 024020, 2011) in a flat FRW universe. The correspondence between scalar 
	field models have been examined by employing new generalized dynamical cosmological term $ \Lambda(t) $. In this regards, the best 
	fit observational values of parameters from three distinct sets data are applied. To decide the solution to field equations, 
	a scale factor $ a= \left(\sinh(\beta t)\right)^{1/n} $ has been considered, where $ \beta$ \& $n $ are constants. Here, we employ 
	the recent ensues ($H_{0}=69.2$ and $q_{0}=-0.52)$ from (OHD+JLA) observation (Yu et al., Astrophys. J. 856, 3, 2018). Through 
	the numerical estimation and graphical assessing of various cosmological parameters, it has been experienced that findings are 
	comparable with kinematics and physical properties of universe and compatible with recent cosmological ensues. The dynamics and 
	potentials of scalar fields are clarified in FRW scenario in the present model. Potentials reconstruction is highly reasonable 
	and shows a periodic establishment and in agreement with latest observations.     
	
	\end{abstract}
	

	
\section{Introduction}

Various experimental observations like SN Ia \cite{ref1}-\cite{ref2}, large scale structure (LSS) \cite{ref3} and Planks data \cite{ref4} etc. 
confirm accelerated expansion of present Universe. Apart from the baryon matter, a large amount of mysterious energy known as dark energy (DE) 
with large critical energy density and huge negative pressure is present in the universe, which is accelerating expansion of the universe. 
Recent observations supported the fact that nearly 70\% matter/energy are in form of DE. The simplest effective component to describe the 
dynamics of the present expanding phase of universe is the Cosmology constant ($ \Lambda $), as it possess a repulsive character. But the 
theoretical models with ($ \Lambda $) faces problems of cosmic coincidence and fine tuning. Therefore, a dynamic cosmological constant 
($ \Lambda = \Lambda(t)$) with negative pressure has been considered a promising candidate of DE. The cosmological constant $ \Lambda $ is considered 
as a variable quantity by different authors for description of expanding universe. Chen {\it et al.} \cite{ref5} took $ \Lambda $ as a 
function of $ \frac{1}{a^2} $, whereas Arbab \cite{ref6} describe $ \Lambda $ as a function of $ \frac{\ddot{a}}{a} $. Waga \cite{ref7} and 
Carvalho {\it et al.}\cite{ref8}  explored $ \Lambda $ in terms of $ (\frac{\dot{a}}{a})^2 $. Recently, several authors develop different 
cosmological models following these particular phenomena, but most of these face certain theoretical and observational restrictions. 
Since observational data do not favour the growth of linear cosmic perturbations, in this way $ \Lambda \propto H^2 $  and 
$ \Lambda \propto \dot{H} $ rules are strongly overrode \cite{ref9}-\cite{ref10}. In case of perfect fluid with equation of state 
$ p = \omega \rho $, cosmological development appears with no DE scenario if $ \Lambda \propto \rho $ is the only assumption taken 
into consideration. The introduction of $ \Lambda \propto \rho $ effectively modifies the proportionality constant $ \Lambda $ and 
the model becomes equivalent to one with merely a new fluid only. We tried to overcome this issue by introducing a new decay law which 
combines both matter and geometry into the dynamical cosmological term.  Many theoretical models like tachyon field, k-essence, quintessence 
scalar field models, quintom, phantom field and chaplygin gas have been proposed recently \cite{ref11}-\cite{ref19}.\\ 

Modified theories of gravitation have been largely focused regarding their cosmological consequences. Several researchers have been 
reviewing DE theories in distinct context of modified speculations of gravitation \cite{ref20}-\cite{ref21}. Despite the sturdy sign 
of the late-time acceleration of universe and also the existence of DE and substance, a great interest are developed in the modified 
theories of gravity. Amongst all such theories, $f(R)$ theory is a suitable theory which probed Einstein's (GR) theory by using the 
arbitrary function $f(R)$ of the gravitational action $R$. $ f(R) $ theory  depicts more developed conditions as compared with GR 
and provides broad measures. Starobinsky \cite{ref22} proposed a singularity free isotropic cosmological model and Sotiriou \cite{ref23} 
gives scalar-tensor theory in frame work of $f(R)$ gravity. Nojiri and Odintsov\cite{ref24} reconstruct the modified theories to get 
cosmological  behaviour of universe and analyzed Big Rip along with the future singularities. The inflation model of $f(R)$ have been 
explored by Huang \cite{ref25}. In the similar manner, numerous cosmologists have also explored $f(T)$ gravity models in distinct context. 
Other modified theory models are $f(R)$ gravity, $f(T)$ gravity, scalar-tensor speculations, Galileon gravity, Braneworld models, 
Gauss-Bonnet gravity etc. \cite{ref26}-\cite{ref32}. Houndjo \cite{ref33} has reconstructed $f(R, T)$ cosmological model showing the 
transition from a decelerated to an accelerated era. Ahmed et al., \cite{ref34} conferred the Bianchi type-V accelerating model 
with $\Lambda(T)$. Some recent researches in $f(R, T)$ gravity have been discussed in \cite{ref34a}-\cite{ref34j} in different contexts.\\

Singh et al. \cite{ref35} have described signature changed model from early decelerated to present accelerated of the universe 
through the bulk viscosity effects in $f(R, T)$ theory of gravity. Sahoo et al. \cite{ref36} established the Kaluza Klein cosmological 
model in $f(R, T)$ theory. Bamba et al. \cite{ref37} explored expansion of universe in presence of DE in f(R) gravity using 
inhomogeneous equation of the state. Sebastiani et al. \cite {ref38} has develop k-essence cosmology in $f(T)$ theory. 
Guendelman et al. \cite {ref39} described a quintessence DE model showing the dynamic effect of vacuum energy density and dust-like matter. 
Aktas¸\cite{ref40} has described various DE models in $f(R, T)$ theory like the k-essence, tachyon field, and quintessence in FRW universe 
with variable $\Lambda$ and $G$ by considering a linearly varying deceleration parameter (DP). The role of $G$ and $\Lambda$ become important 
in explaining accelerating expansion of the universe. Various authors have consider 
the cosmological constant and $G$ as time function in numerous frameworks. Norman Cruz\cite{ref41} assume a relationship between the DE 
densities of k-essence and holographic, in this case the HDE is defined in form of k-essence scalar field for $c > 1$. Granda et al. \cite{ref42} 
formulate a relationship between energy density of k-essence, dilation, quintessence, and tachyon with HDE in the flat FRW universe. 
This correlation permits the models of the scalar fields to reconstruct the potentials and dynamics to characterize accelerated expansion. 
A general Lagrangian density $p(\phi,X)$ scalar-field energy model was reconstructed by Tsujikawa \cite{ref43}. Korumur \cite{ref44} and 
Sharif and Jawad \cite{ref45} developed DE models using scalar field in the flat Kaluza-Klein universe. The most attractive 
and exciting state in cosmology is the detection of the accelerated expansion of the universe \cite{ref46}-\cite{ref49}. So many ways are 
there to address this problem, including: models of k-essence, the quintessence field, the cosmological constant, a brane and cosmology 
scenarios \cite{ref50}-\cite{ref57}. \\

We have revisited in the recent work \cite{ref58} by considering a distinct value of the cosmological term $\Lambda$ as a simple linear 
combination of energy density $\rho$ and scalr factor $a$ and obtained a new and better scenario for different dark energy.
The scalar field DE models can effectively clarify the basic DE theory, the application of scalar field DE models become very exciting 
to describe the energy density as influential theory. In the present work, we discuss the evaluation of the scalar field with the tachyon, 
k-essence, and the quintessence models in the framework of the $f(R, T)$ gravity. We consider a new generalized dynamic form of $\Lambda$. 
Sect. $2$ consists of field equations. In Sect. $3$, we propose the solution of the field equations. Sect. $4$ contains three sub-sections 
expressing the tachyon, k-essence, and quintessences scalar fields. Sect. $5$ consists concluding summary and results of derived model.
	
\section{ Field Equations in f(R, T) gravity}

We consider the modified $ f(R,T) $ theory first researched by Harko et al.\cite{ref59}: 

\begin{equation}
\label{1}
S=\int \left(\frac{\sqrt{-g}}{2 k} f(R,T)+\sqrt{-g} L_{m}\right)dx^4,
\end{equation}

 The Einstein's field equations with variable $ \Lambda $ and $ G $ in the framework of $ f(R,T) $ theory are assumed \cite{ref60}:

\begin{equation}
\label{2}
G_{\alpha \beta}=\left[8 \pi G(t)+2\lambda'_{a}(T)\right]T_{\alpha \beta}+\left[2 p \lambda'_{a}(T) +\lambda_{a}(T)+\Lambda(t)\right]g_{\alpha \beta},
\end{equation}

Here `dash' represent the differentiation.\\
By assuming $\lambda_{a}(T)=\mu T $, Eq.(\ref{2}) can be rewrite as  \cite{ref60},
\begin{equation}
\label{3}
G_{\alpha \beta}=\left[8 \pi G(t)+2\mu \right]T_{\alpha \beta}+\left[\mu \rho-\mu  p +\Lambda(t)\right]g_{\alpha \beta},
\end{equation}
where ( $\mu  \rightarrow $ constant).\\

 Now, we will report about the distinct DE models in the formalism of $ f(R,T) $ theory taking variable $ G $ \& $ \Lambda $ 
 in FRW space-time. By applying $ \mu=0 $ in Eq. (\ref{3}), we obtained GRT scenario.\\
 
 The spatially flat FRW universe is given by

\begin{equation}
\label{4}
ds^{2}=-dt^{2}+a^{2}(t)\left(dx^{2}+dy^{2}+dz^2\right).
\end{equation}

 the Ricci scalar is given as:

\begin{equation}
\label{5}
R=-6\left(\frac{\ddot{a}}{a}+\frac{\dot{a}^2}{a^2}\right)
\end{equation}

 Here, the energy-momentum tensor can be determined as:

\begin{equation}
\label{6}
T_{\alpha\beta}= -p g_{\alpha\beta}+(\rho+p)u_{\alpha}u_{\beta},
\end{equation}

where $ \rho$ and $ p$ stand for energy density and cosmic pressure respectively.
$ u^{i}=(0,0,0,1) $ indicates the four-velocity components. The trace of energy-momentum tensor is given by $ T^{DF}=\rho-3p $.  \\

For the flat FRW universe, the field equations of DE model in framework of $ f(R,T)$ theory are described as:

\begin{equation}
\label{7}
3 H^2+2 \dot{H}=-8\pi G p +\mu \rho-3\mu p+\Lambda
\end{equation} 

\begin{equation}
\label{8}
3 H^2-8\pi G \rho=3\mu \rho+\Lambda-\mu p
\end{equation}  

Here, $ H=\frac{\dot{a}}{a} \to$ Hubble parameter.

\section{ Solution of the Field Equations}
We have two independent field equations (\ref{7}) and (\ref{8}) in five unknowns $H$, $\rho$, $p$, $G$, and $\Lambda$.
 Thus, to determine explicit solution of field  equations, we need three more assumptions 
 consider the some more assumptions as:

\begin{itemize}
\item[(i)] We assume the dynamic cosmological term $ \Lambda (t) $ as a simple linear combination of $ \rho $ and $ a $ \cite{ref61}: 
\[ \Lambda=l \frac{\ddot{a}}{a}+\lambda \left(\frac{\dot{a}}{a}\right)^2+4 \pi G \delta \rho \]
where $ l $, $ \lambda $ and $ \delta $ are constants. 
\item[(ii)] The EoS parameter is consider as $ \omega=\frac{p}{\rho} $.
\item[(iii)] A deceleration parameter is supposed as \cite{ref62a,ref62b,ref62c,ref62d}
\begin{equation}/
\label{9}
q=-\frac{a \ddot{a}}{\dot{a}^2}=n\left[-\tanh^2{(\beta t)} +1\right]-1
\end{equation} 
\end{itemize}

Using above assumption in Eqs. (\ref{7}), Eq. (\ref{8}), the expressions for physical quantities $\rho$, $p$, $G$, and $\Lambda$ are derived as:

\begin{equation}
\label{10}
\rho=\frac{\beta ^2 \text{csch}^2(\beta  t) \left[(\omega+1)\left\{(\lambda-3)-l(2n-1)+(\lambda+l-3)\cosh (2 \beta  t)\right\}+
2n(\delta+2)\right]}{2 \mu  n^2 (\omega+1) (\delta +\omega-1)}
\end{equation}  
\begin{equation}
\label{11}
p=\frac{\omega\beta ^2 \text{csch}^2(\beta  t) \left[(\omega+1)\left\{(\lambda-3)-l(2n-1)+(\lambda+l-3)\cosh (2 \beta  t)\right\}+
2n(\delta+2)\right]}{2 \mu  n^2 (\omega+1) (\delta +\omega-1)}
\end{equation} 

\begin{equation}
\label{12}
G=-\frac{\mu  \left[(\omega+1) \left\{(\lambda-3)-l(2n-1)+(\lambda+l-3)\cosh (2 \beta  t)\right\}-2n(\omega-3)\right]}{4 \pi \left[(\omega+1) 
\left\{(\lambda-3)-l(2n-1)+(\lambda+l-3)\cosh (2 \beta  t)\right\}+2n(\delta+2)\right]}
\end{equation}

\begin{eqnarray}
\label{13}
&\Lambda=\frac{\beta^2}{n}\left[\frac{\delta (\omega-3)}{(\omega+1) (\delta +\omega-1)}-l\right] \text{csch}^2(\beta  t)+
\frac{\beta^2 (\lambda +l)}{ n^2}  \coth ^2(\beta  t)\nonumber\\
&-\frac{\beta^2 \delta}{2 n^2} \left[\frac{ (\lambda-3)-l(2n-1)+(\lambda+l-3)\cosh (2 \beta  t)}{(\delta +\omega-1)}\right] 
\text{csch}^2(\beta  t).
\end{eqnarray}  

From Eq. (\ref{5}), Eq. (\ref{6}), the trace and Ricci scalar of DE matter for $f(R, T) = R+2 \mu T$ model are determined  as:

\begin{equation}
\label{14}
R=6 \left(\frac{n \beta^2-2\beta^2 \cosh^{2}{(\beta t)}}{n^2 \sinh^{2}{(\beta t)}}\right)
\end{equation}  

\begin{equation}
\label{15}
T^{DF}=(1-3\omega)\left(\frac{(\mu-3)(1+\omega) \beta^2 \cosh^2{(\beta t)}+\beta^2 n}{n^2 \lambda (\omega^2-1) \sinh^2{(\beta t)}}\right)
\end{equation}  
Using Eq. (\ref{14}) and Eq. (\ref{15}), we get $ f(R,T)=R+2\mu T $ as
\begin{equation}
\label{16}
f(R,T)=6 \left(\frac{n \beta^2-2\beta^2 \cosh^{2}{(\beta t)}}{n^2 \sinh^{2}{(\beta t)}}\right)+2(1-3\omega)
\left(\frac{(\mu-3)(1+\omega) \beta^2 \cosh^2{(\beta t)}+\beta^2 n}{n^2 (\omega^2-1) \sinh^2{(\beta t)}}\right)
\end{equation} 

From Eq(\ref{9}), it has been found that $ q>0 $ for $ t<\frac{1}{\beta}\tanh^{-1}\left(1-\frac{1}{n}\right)^{1/2} $ and 
$ q<0 $ for $ t>\frac{1}{\beta}\tanh^{-1}\left(1-\frac{1}{n}\right)^{1/2} $. Our model is also shown to be in an accelerating phase 
for $ 0<n\leq 1 $, but the universe transitions from a decelerating to an accelerating phase for $ n\geq 1 $. Values of $ n $ and $ \beta $ 
can be used in the resulting model to ensure that the decelerating parameter behaves properly. The current universe is accelerating, according 
to recent SNe Ia observations, and the value of DP is in the range $ -1 \leq q < 0 $. \\

The DP in terms of red shift $ z $ is given by
\begin{equation}
	\label{17}
	q(z)=-1+n+\frac{n (1-n+q_{0})}{n+1+q_{0}[-1+(1+z)^{2n}]}
\end{equation}
where $ z=-1+\frac{a_{0}}{a} $ is red shift parameter, $ a_{0} $ is the present value of scale factor, $ q_{0} $ is present value of DP at $ z=0 $.

The metric potential for the model form Eq. (\ref{9}) is obtained as,  
\begin{equation}
	\label{18}
a=\left[\sinh{(\beta t)}\right]^{1/n}
\end{equation}
where $ n $ and $ \beta $ are constants. 

From Eq. (\ref{9}), a relationship between $n$ and $\beta$ can be considered for the present universe ($ t_{0}=12.36 $ Gyr with 
$q_{0} = -0.52 $ Yu et al. \cite{ref63}) as.

\begin{equation}
	\label{19}
	\beta=\frac{1}{12.36} \tanh^{-1}\left[1-\frac{0.48}{n}\right]^{1/2}
\end{equation}
which states that the model is valid for $ 0.48 < n \leq 1 $ and shows accelerated expansion of present universe.

From Eq.(\ref{19}), we get a set of values ($ n=2 $, $ \beta=0.1084504 $), these experimental values of $ n $ and $ \beta $ can be used 
for plotting and validation of the given model.\\

Now, we consider the Hubble data sets $ H(z) $ \cite{ref64}-\cite{ref69}, 580 data sets of apparent magnitude $ m(z) $ from union 2.1 compilation 
of SNe Ia data sets \cite{ref70} and 51 data sets of $ m(z) $ from Joint Light Curve Analysis (JLA) data sets \cite{ref71} and using the 
$R^2$-test formula, we obtain the best fit values of various model parameters $ l, \lambda $ and $ \delta $ for the best fit curve of Hubble 
function H(z) and apparent magnitude m(z). The best fit values of $ l, \lambda $ and $ \delta $ for the various data sets  at 95\% level of 
confidence are mentioned in the table-1. These values of parameters are used for plotting and describing the dynamics of model.

\begin{table}[htb] 
	\centering
	\begin{tabular}{|c|c|c|c|}
		\hline
		Parameter  & H(z)  & SNe Ia & JLA  \\ [1ex]
		\hline
		$ l $ & 1.764 & 0.1881 & 0.1137  \\
		\hline
		$ \lambda $ & 0.7351 & 0.8632 & 0.9196 \\
		\hline
		$ \delta $ & 0.4916 & 0.7292 & 0.9696  \\
		\hline
	\end{tabular}
	\caption{ Table of values of parameters} 
	\label{table1} 
\end{table}

\begin{figure}[t!]
	\centering
	\includegraphics[scale=1]{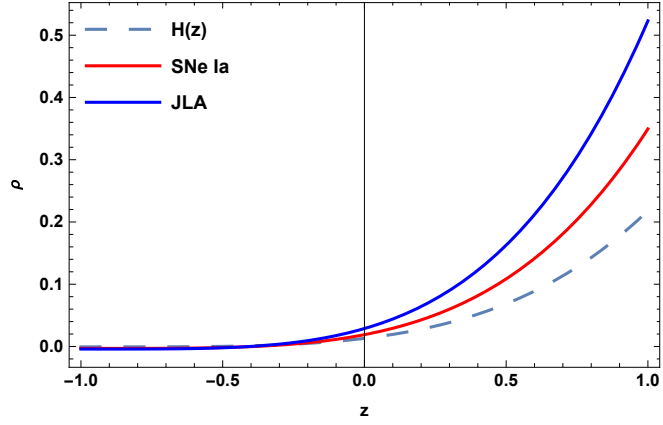}
	\caption{Plot of Energy density $ \rho $ against $ z $ with $n=2, \beta=0.1084504,\mu=-2, \omega=-2/3 $.}\label{fig1}	
\end{figure}
\begin{figure}[t!]
	\centering
	\includegraphics[scale=1]{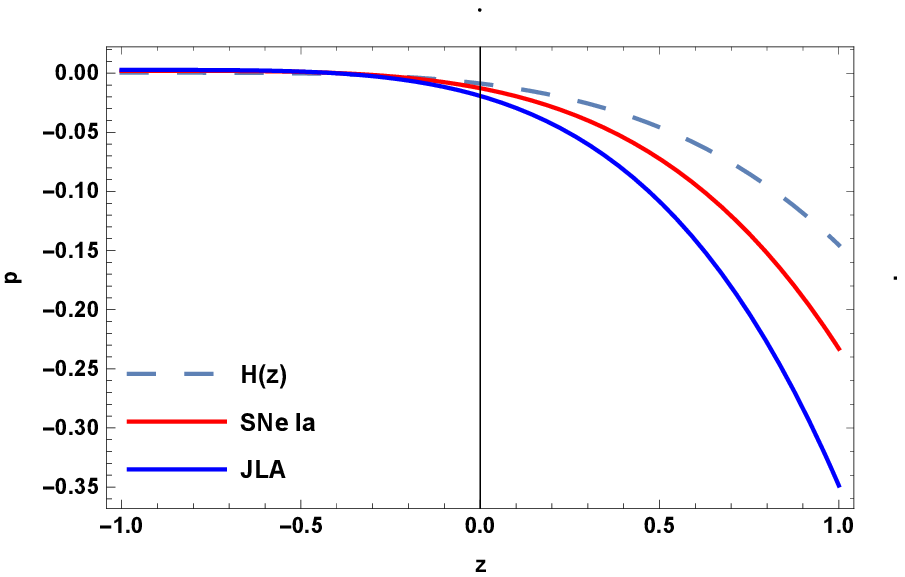}
	\caption{Variation of pressure $ p $ versus $ z $ with $ n=2, \beta=0.1084504,\mu=-2, \omega=-2/3 $. }\label{fig2}	
\end{figure}
\begin{figure}[t!]
	\centering
	\includegraphics[scale=1]{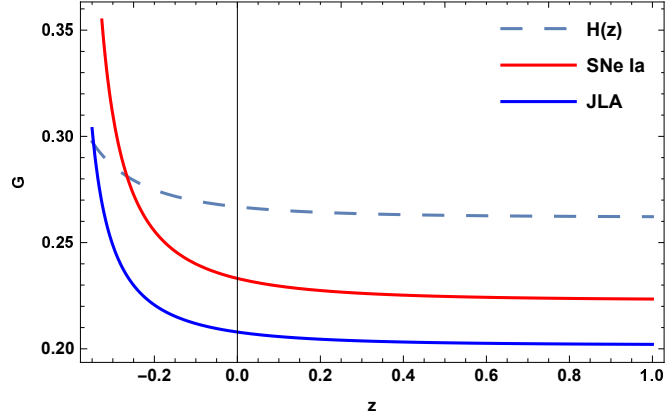}
	\caption{Plot of gravitational term $ G $ versus  $z $ with $n=2, \beta=0.1084504,\mu=-2, \omega=-2/3 $.}\label{fig3}	
\end{figure}
\begin{figure}[t!]
	\centering
	\includegraphics[scale=1]{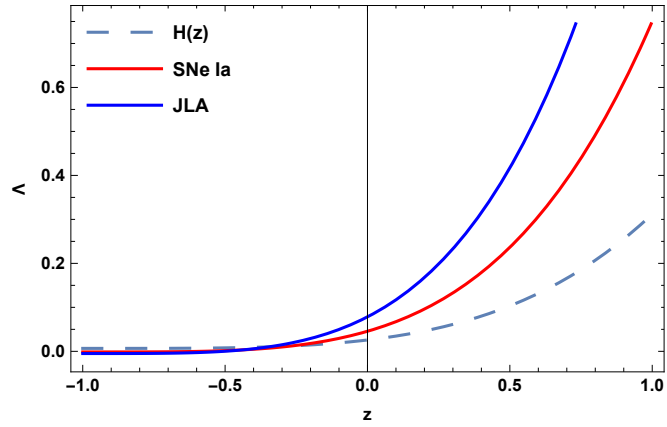}
	\caption{Plot of cosmological term $ \Lambda $ versus $ z $ with $n=2, \beta=0.1084504,\mu=-2, \omega=-2/3 $.}\label{fig4}	
\end{figure}

Figure $1$, depict the variation of  $\rho$ against red shift $z$ for sets of observational data. 
Figure shows that $\rho$ is a positive decreasing function that approaches to zero in the current era. 
It's worth noting that the energy density $\rho$ decrease sharply for 
$H(z)$ data as compared to SNe Ia and JLA observational data.\\

Figure $2$ shows the variability of fluid pressure $p$ against the redshift $z$ for three observations data sets. Figure confirms that the 
isotropic pressure is negative all over the expansion. It has been observed from the figure that pressure is a increasing function and 
converge to zero in late time. The high negative pressure corresponding to JLA observational data in comparison with set of observation 
data explains the accelerated expansion of the universe.\\

In the late time the gravitational term $G(t)$ rises slowly and leads to infinity as noticed from the Figure $3$. It is clearly 
observed from the figure that $G$ increases correspond to all set of observations but for SNe Ia data, it increase very sharply. \\

The cosmological term $\Lambda$ explicate the behaviour of the universe in the present cosmological model. Figure $4$ clarify the nature of 
$\Lambda$ against redshift $z$. One can easily observed from the figure that  $\Lambda$ is a decreasing function and approach to zero in 
present epoch. Our universe is in accelerated expansion era as specified by various recent studies \cite{ref72,ref73}. In this regards, 
behaviour of $\Lambda$ in the derived model is in good agreement with observations and trends. However, as compared to $H(z)$ and SNe Ia 
observational data sets, it decreases for JLA universe.\\

 
\section{The f(R, T) DE Models with correspondence of  scalar fields}

The scalar field DE models are efficient to explain the occurrence of DE.  There are various such model already mentioned in the literature 
like quintessence, condensate, k-essence, ghost dilaton phantom and tachyon, etc. \\

Reconstruction of the efficient DE models necessarily explain the nature of quantum gravity by the means of scalar fields. These models 
demonstrate quintessence nature of the universe and generate an efficient explanation of the existence of DE in the universe. In this section, 
we search the DE models in correspondence with tachyon, k-essence and, quintessence scalar fields taking variable $G$ and $\Lambda$ in the 
frame work of $f(R, T)$ theory. In derived model, we assume the values of EoS parameter as $ \omega=-1/3$ and $-2/3$. Now, the behaviour of 
kinetic energy and scalar potential against $z$ are demonstrated through the plots for three sets of observational data by employing 
$a= a_{0}(1+z)^{-1}$, here we assume $ a_{0}=1$ as considered by distinct  authors \cite{ref58,ref74}.

\subsection{DE model with Tachyon field }


The tachyon field (TF) is one of the dark energy components used to characterize the universe's rapid expansion \cite{ref75,ref76}.
Some tachyon condensate with an effective Lagrangian density \cite{ref77,ref78} suggests a tachyon field.\\

TF is considered as one of the DE constituent which explain the accelerated expansion of the universe. The EoS parameter of the tachyon DE 
matter distribution falls between $-1$ and $0$ Gibbon \cite{ref79}. For tachyon matter distribution, the energy density $ \rho $ and  pressure $ p $ 
associated to SF ($ \Phi $) and scalar potential $ V(\Phi)$ in flat FRW context are as follows:

\begin{equation}
\label{20}
p_{TF}=-T^{i}_{i}=V(\Phi) \left(1-\dot{\Phi}^2\right)^{1/2}
\end{equation}

\begin{equation}
\label{21}
\rho_{TF}=T^{4}_{4}=\frac{V(\Phi)}{\sqrt{1-\dot{\Phi}^2}} 
\end{equation}

Here, $ \dot{\Phi}^2 \to$ kinetic energy (KE),  $ V(\Phi) \to$  scalar potential.\\
 $\Phi$ and $V(\Phi)$ in terms of redshift are obtained as:


\begin{figure}[t!]
	\centering
	\includegraphics[scale=0.80]{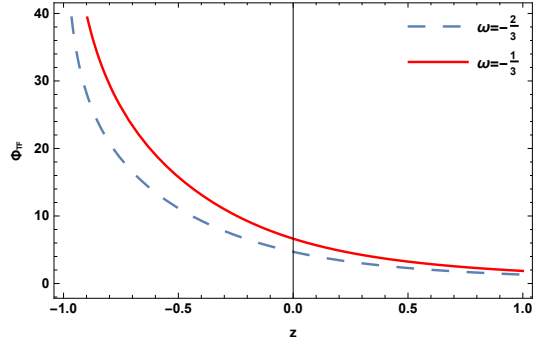}
	\caption{Plot of SF for tachyon against $ z $  with $n=2.0, \beta=0.1084504$ and $ c_{1}=0 $. }\label{fig5}	
\end{figure}

\begin{figure*}[t!]
	\centering
	\begin{subfigure}[t]{0.45\linewidth}
		\centering
		\includegraphics[width=\linewidth]{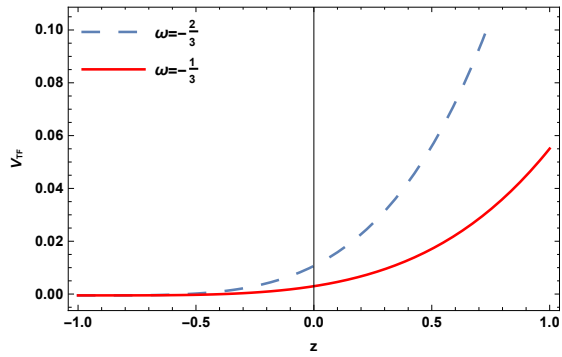}
		\caption{$l=1.764,\lambda=0.7351,\delta=0.4916$, (H(z) data)} 
	\end{subfigure} 
	\quad\quad\begin{subfigure}[t]{0.45\linewidth}
		\centering
		\includegraphics[width=\linewidth]{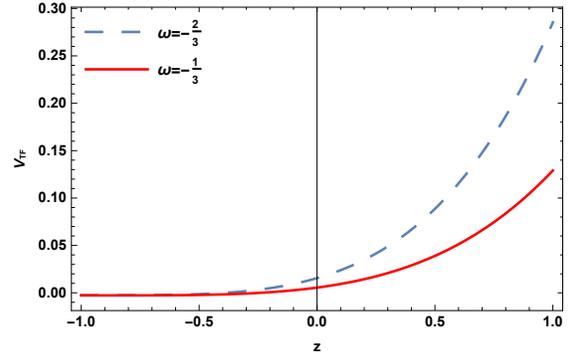}
		\caption{$l=0.1881,\lambda=0.8632,\delta=0.7292$ (SNe Ia data)}
	\end{subfigure}\\
	\quad\quad\begin{subfigure}[t]{0.45\linewidth}
		\centering
		\includegraphics[width=\linewidth]{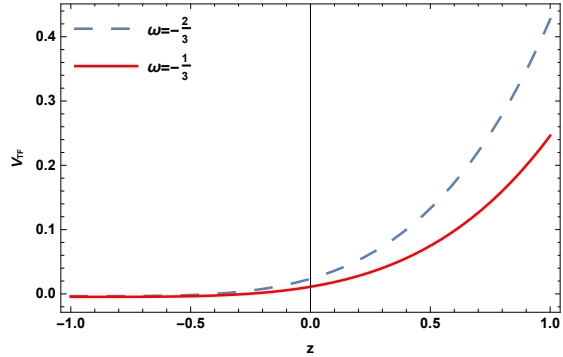}
		\caption{$l=0.1137,\lambda=0.9196,\delta=0.9696$ (JLA data)}
	\end{subfigure}
	\caption{Plot of $ V(\Phi) $ for tachyon field against $ z $  with $ n=2.0, \beta=0.1084504,\mu=-2$.}\label{fig6}
\end{figure*}
\begin{equation}
\label{22}
\Phi_{TF}=\frac{\left(\omega+1\right)^{1/2} \sinh ^{-1}\left(\left({z+1}\right)^{-n}\right)}{\beta }+c_{1}
\end{equation}  

\begin{equation}
\label{23}
V(z)_{TF}=\frac{\beta ^2 \sqrt{-\omega} \left(\frac{1}{z+1}\right)^{-2 n} \left(l \left(\left({z+1}\right)^{-2n}+1-n\right)+
\frac{(\delta +2) n}{\omega+1}+(\lambda -3) \left(\left({z+1}\right)^{-2n}+1\right)\right)}{\mu  n^2 (\delta +\omega-1)}.
\end{equation}

Figure $5$, represents the variation of the kinetic tachyon energy with respect to redshift $z$ for observational values \cite{ref63} 
by assuming the $\omega = -1/3~ $ \& $~ -2/3$.  Plots shows that kinetic energy is increasing along the redshift $z$. The SF ($ \Phi $) 
is very low in early universe while it is very high in present era.\\

Similarly, Figures 6(a,b, \& c) depict the effect scalar potential against $(z)$. $V(\Phi)$ against $ z $ decreases for all the three set of 
observations. From the above discussion for the tachyon potential, we observe that the tachyon field decreases as the universe expands 
\cite{ref63}-\cite{ref70}.

\subsection{DE model with k-essence field }
	 
To explain the universe's late-time acceleration, the k-essence scalar field (SF) model was proposed which is motivated by the Born-Infeld 
action string theory \cite{ref80,ref81}. \\ 

k-essence models produced a very attractive dynamical scenario of the universe, which shows th nature in agreement with current cosmological 
observations. Thus, the fine tuning of the initial scalar field conditions can be ignored in these models\cite{ref82}. 
In a flat FRW background, $p$ and $\rho$ in terms of $ \Phi $ and $ V(\Phi) $ for k-essence field are established as \cite{ref83}: 

\begin{equation}
\label{24}
p_{ke}=-T^{i}_{i}=\frac{V(\Phi) \left(\dot{\Phi}^4-2 \dot{\Phi}^2\right)}{4}
\end{equation}  

\begin{equation}
\label{25}
\rho_{ke}=T^{4}_{4}=\frac{V(\Phi) \left(3\dot{\Phi}^4-2 \dot{\Phi}^2\right)}{4}
\end{equation}

The $\Phi$ and $V(\Phi)$ for the k-essence field are establish as:

\begin{equation}
\label{26}
\Phi_{ke}=\frac{\sqrt{\frac{2 \omega -2}{3 \omega -1}} \sinh ^{-1}\left(\left(\frac{1}{z+1}\right)^n\right)}{\beta }+c_{2}
\end{equation}  
\begin{equation}
\label{27}
V(z)_{ke}=\frac{\beta ^2 (3 \omega-1)^2 \left(\frac{1}{z+1}\right)^{-2 n} \left(l \left(\left(z+1\right)^{-2 n}+1-n\right)+
\frac{(\delta +2) n}{\omega+1}+(\lambda -3) \left(\left(z+1\right)^{-2 n}+1\right)\right)}{2 \mu  n^2 (1-\omega) (\delta +\omega-1)}
\end{equation}

\begin{figure}[t!]
	\centering
	\includegraphics[scale=0.80]{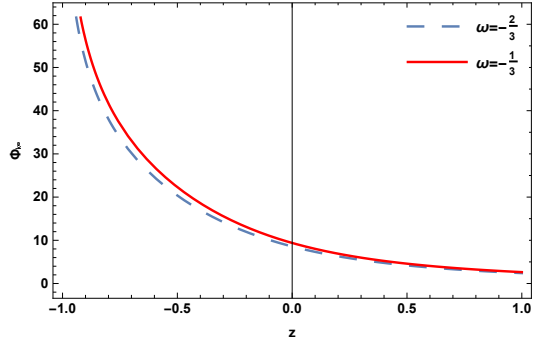}
	\caption{Plot of SF for k-essence against $ z $  with $ n=2.0, \beta=0.1084504$ and $ c_{2}=0 $. }\label{fig7}	
\end{figure}
\begin{figure*}[t!]
	\centering
	\begin{subfigure}[t]{0.45\linewidth}
		\centering
		\includegraphics[width=\linewidth]{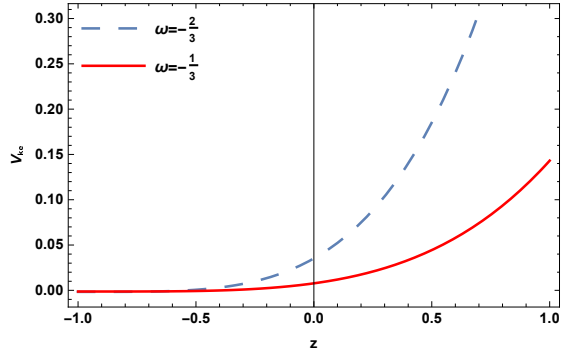}
		\caption{$l=1.764,\lambda=0.7351,\delta=0.4916$, (H(z) data)} 
	\end{subfigure} 
	\quad\quad\begin{subfigure}[t]{0.45\linewidth}
		\centering
		\includegraphics[width=\linewidth]{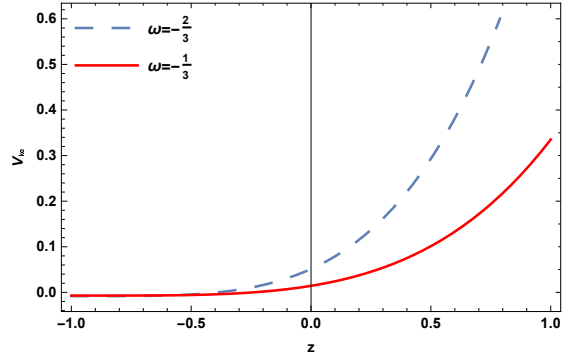}
		\caption{$l=0.1881,\lambda=0.8632,\delta=0.7292$ (SNe Ia data)}
	\end{subfigure}\\
	\quad\quad\begin{subfigure}[t]{0.45\linewidth}
		\centering
		\includegraphics[width=\linewidth]{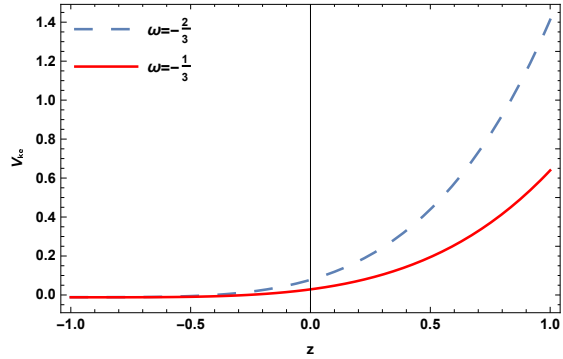}
		\caption{$l=0.1137,\lambda=0.9196,\delta=0.9696$ (JLA data)}
	\end{subfigure}
	\caption{Plot of $ V(\Phi)$ for k-essence against $ z $  with $  n=2.0, \beta=0.1084504,\mu=-2 $.}\label{fig8}
\end{figure*}

Plot $7$, demonstrate the k-essence KE with $z$ for recent observations \cite{ref63} by choosing $\omega = -1/3 ~ $\& $ ~-2/3$. The graph 
convey that the KE is very high in late universe and express a behaviour similar to tachyon. Figure 8(a,b,c), draw the scalar potential 
effect against $(z)$, it decreases with redshift and approach to zero in late universe for three sets of observations. For $\omega =-1$, 
derived model shows a singularity \cite{ref63}-\cite{ref70}.

\subsection{DE model with Quintessence field}


\begin{figure}[t!]
	\centering
	\includegraphics[scale=0.60]{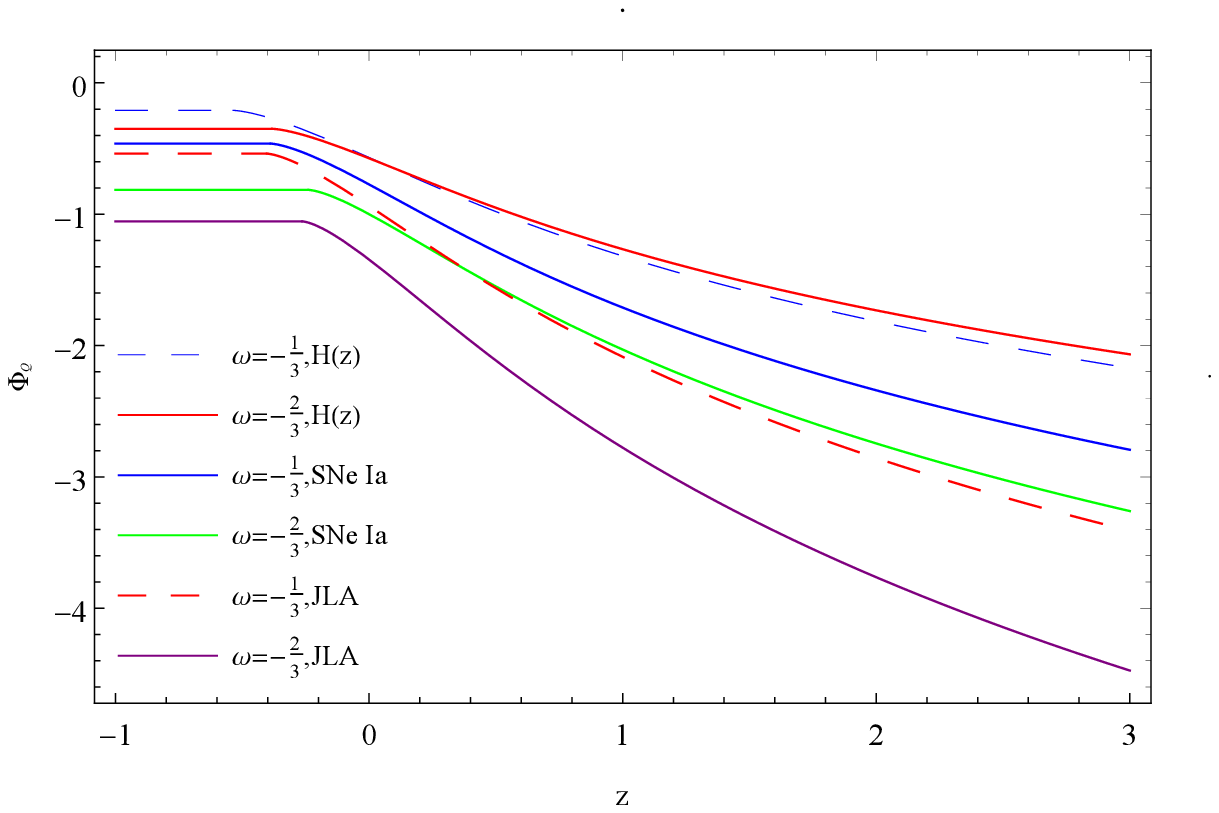}
	\caption{Plot of $ \Phi $ for Quintessence against $ z $ with $  n=2.0, \beta=0.1084504,\mu=-2 $ and $ c_{3}=0 $. }\label{fig9}	
\end{figure}

Quintessence is a theoretical form of DE characterized by homogeneous variable SF as well as the scalar potential, which is responsible 
for universe's expansion \cite{ref81}. For quintessence model, EoS parameter specifies that the accelerated expansion of universe exists 
for $-1\leq\omega \leq -\frac{1}{3}$ \cite{ref81}. The relations of $ \rho $ and $ p $ in form of $ \Phi $ and $ V(\Phi) $ for quintessence 
model in flat FRW universe are established as \cite{ref51}:

\begin{equation}
\label{28}
p_{Q}=-T^{i}_{i}=\frac{1}{2}\left(\dot{\Phi}^2-2V(\Phi)\right)
\end{equation}  

\begin{equation}
\label{29}
\rho_{Q}=T^{4}_{4}=\frac{1}{2}\left(\dot{\Phi}^2+2V(\Phi)\right)
\end{equation}

\begin{figure*}[t!]
	\centering
	\begin{subfigure}[t]{0.45\linewidth}
		\centering
		\includegraphics[width=\linewidth]{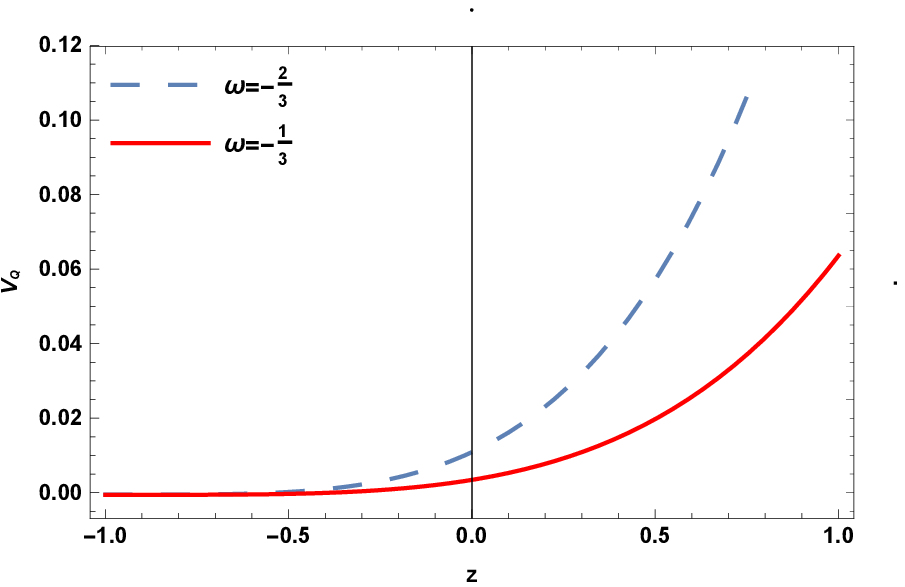}
		\caption{$l=1.764,\lambda=0.7351,\delta=0.4916$, (H(z) data)} 
	\end{subfigure} 
	\quad\quad\begin{subfigure}[t]{0.45\linewidth}
		\centering
		\includegraphics[width=\linewidth]{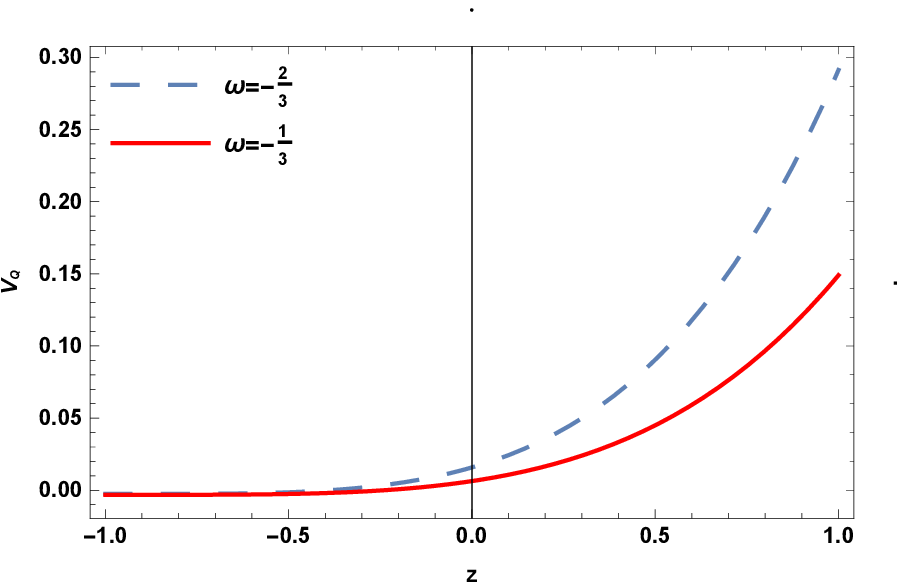}
		\caption{$l=0.1881,\lambda=0.8632,\delta=0.7292$ (SNe Ia data)}
	\end{subfigure}\\
	\quad\quad\begin{subfigure}[t]{0.45\linewidth}
		\centering
		\includegraphics[width=\linewidth]{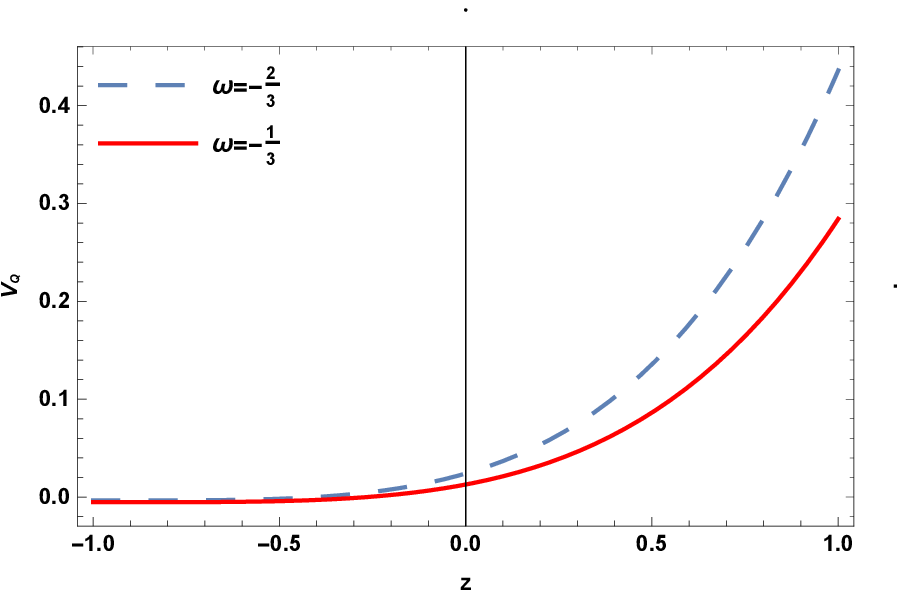}
		\caption{$l=0.1137,\lambda=0.9196,\delta=0.9696$ (JLA data)}
	\end{subfigure}
	\caption{Plot of $ V(\Phi) $ for quintessence against $ z $  with $  n=2.0, \beta=0.1084504,\mu=-2 $.}\label{fig10}
\end{figure*}


The $\Phi$ and $V(\Phi)$ for the quintessence model are determined as:
\begin{eqnarray}
\label{30}
&\Phi_{Q}=\frac{a_{2}}{a_4}
 \tanh ^{-1}\left(\sqrt{2}\frac{a_{2}}{a_{1}} \cosh \left(\sinh ^{-1}\left(\frac{1}{z+1}\right)^n\right)\right)\nonumber\\
&-\frac{a_{3}}{a_4} \log\left(\sqrt{2} a_{3} \cosh \left(2 \sinh ^{-1}\left(\frac{1}{z+1}\right)^n\right)+a_{1}\right)+c_{3}
\end{eqnarray}  
where, 
\[ a_{1}=\sqrt{(\omega+1) \left\{(\lambda-3)-l(2 n-1)+(\lambda +l-3) \cosh \left(2 \sinh ^{-1}
\left(\frac{1}{z+1}\right)^n\right)\right\}+2 n(\delta+2)}\] 

\[ a_{2}=\sqrt{-l (n-1) (\omega+1)+(\delta +2) n+(\lambda -3) (\omega+1)} \]

\[a_{3}=\sqrt{(\omega+1) (\lambda +l-3)}\] 

\[ a_{4}=\sqrt{\mu  n^2 (\delta +\omega-1)} \]

\begin{equation}
\label{31}
V(z)_{Q}=\frac{\beta ^2 (1-\omega) \left(\frac{1}{z+1}\right)^{-2 n} \left(l \left(\left(\frac{1}{z+1}\right)^{2 n}-n+1\right)+
\frac{(\delta +2) n}{\omega+1}+(\lambda -3) \left(\left(\frac{1}{z+1}\right)^{2 n}+1\right)\right)}{2 \mu  n^2 (\delta +\omega-1)}
\end{equation}

Plot 9, convey the variation of quintessence KE versus $z$ for set of observations \cite{ref63}-\cite{ref70} by choosing  
$\omega = -1/3 ~ $ \& $ ~-2/3$. Similar to k-essence and tachyon field, the behaviour of the quintessence field is increasing and 
in late time it approaches to zero. Figure 10(a,b,c) depicts the effect of Scalar potential along $(z)$, it decreases and approaches 
to zero in late universe for all three sets of observations. For $\omega =-1$  model shows a kind of singularity.


\section{Concluding Remarks:}

In the present work, cosmological consequences are developed using the scalar fields in the frame work of $f(R, T) $ theory. 
We look at a specific dynamical form of cosmological term $ \Lambda(t)=\Lambda=l \frac{\ddot{a}}{a}+\lambda 
\left(\frac{\dot{a}}{a}\right)^2+4 \pi G \delta \rho $, that has been generated using modification of Einstein-Hilbert action 
and is used to describe cosmological inflation, late-time acceleration of the cosmos, or dark matter in the sense 
that a set of specific scalar field models. The best fit values of $ l, \lambda $ and $ \delta $ for the various data sets  at $95\%$ level 
of confidence for different sets of observational data and are used for plotting the dynamics of model. In this work, we have 
searched various scalar field DE models in flat FRW universe  for with variable $G$ and $\Lambda$ in formalism of $ f(R,T) $ 
theory by considering different set of cosmological observational values.\\

The following are the primary outputs of the generated model: \\
\begin{itemize} 
\item In the present model, Fig. $1$ shows that the energy density $\rho$ is a positive decreasing function and approaches to zero 
in present epoch, which is consistent with the Big bag theory. It is important to note that energy density $\rho$ decrease sharply 
for $H(z)$ data as compared to SNe Ia and JLA observational data.
 
\item Figure $2$ demonstrate that the pressure $p$ is negative throughout the cosmic era and converge to zero in late time. It has been 
observed from the figure that the high negative pressure exists corresponding to JLA observational data in comparison with other sets of 
observations. The rapid expansion of the cosmos is caused by this negative pressure. 
 
\item $G(t)$ increases slowly and tends to infinity in the late universe as noticed from the Figure $3$. It is clearly observed from the 
figure that the gravitational term increases correspond to all set of observations but for SNe Ia data, it increase very sharply. 
The cosmological constant $\Lambda$ is a positive constant and has a magnitude of $\Lambda(Gh / c^{3})=10^{-23}$ as shown by 
Schmidt et al. \cite{ref83}. In the derived model, the Figure $4$ shows that  $\Lambda$ is a decreasing function and approach to zero in 
present epoch. Through the findings of the magnitude and redshift of  cosmological $\Lambda$-term by Ia supernova observations, it get 
impression that expansion of universe can be accelerated with induced cosmological density. 
 
\item Scalar field DE models like tachyon, k-essence and quintessence are explored for constant values of EoS parameter $\omega= -1/3, -2/3$ 
as shown in Figs.$5$-$10$. The kinetic energy $\Phi$ and corresponding potential are derived for respective models. We observe that kinetic 
energy $\Phi$ is very high at present epoch while the respective potentials are very low (approach to zero). The derived model shows a kind 
of singularity for k-essence and tachyon scalar fields at $\omega =-1$.  

\item In the end, we can conclude that the present model initiate with Big-Bang scenario while terminate with Big Rip. These 
kinds of results are the consequences of the substantial swing in the evolutionary growth of matter and DE dominated eras. 
The potentials $V(\Phi)$ remains positive for negatively  varying $\mu$. We may derive a GRT solution for FRW world with the tachyon, 
k-essence, and quintessence scalar field distributions in this way since $G = 0$ for $\mu = 0 $. The current model's results describe 
that the current state of universe, which is consistent with the scalar field and its related potential. The formulation dynamics of 
scalar field models are used to explain the tachyon, k-essence, and quintessence cosmologies in the framework of gravity $f(R,T)$. 

\item In previous work the authors \cite{ref58} have taken the ratio between $ H^2 $ and $ \Lambda $, i.e.$ \Lambda=\xi H^2 $, where $ \xi $ is 
a constant. But in the present paper we have considered the dynamic cosmological term $ \Lambda (t) $ as a simple linear combination of $ \rho $ and $ a $  
\[ \Lambda=l \frac{\ddot{a}}{a}+\lambda \left(\frac{\dot{a}}{a}\right)^2+4 \pi G \delta \rho \].

\item In previous work the authors \cite{ref58} assumed the deceleration parameter as:
$q=-\left(\frac{\dot{H}+H^2}{{H}^2}\right)=\frac{m }{(m+kt)^2}-1$, whereas in the present article we consider 
\[q=-\frac{a \ddot{a}}{\dot{a}^2}=n\left[-\tanh^2{(\beta t)} +1\right]-1 \].

\end{itemize}
Hence, the results obtained in the present paper is more general than the previous one \cite{ref58}.
The results in the given work may be useful to understand the evolution of universe through the tachyon, k-essence, and quintessence 
field models with FRW space-time in the formalism of $f(R,T)$ gravity.




\end{document}